# Classifying superconductivity in compressed H$_3$S

E. F. Talantsev[1,2]

[1]M.N. Mikheev Institute of Metal Physics, Ural Branch, Russian Academy of Sciences, 18, S. Kovalevskoy St., Ekaterinburg, 620108, Russia

[2]NANOTECH Centre, Ural Federal University, 19 Mira St., Ekaterinburg, 620002, Russia

E-mail: evgeny.talantsev@imp.uran.ru

*Abstract*

The discovery of high-temperature superconductivity in compressed H$_3$S by Drozdov and co-workers (A. Drozdov, et. al., *Nature* **525**, 73 (2015)) heralded a new era in superconductivity. To date, the record transition temperature of $T_c$ = 260 K stands with another hydrogen-rich compound, LaH$_{10}$ (M. Somayazulu, et. al., arXiv:1808.07695) which becomes superconducting at pressure of $P$ = 190 GPa. Despite very intensive first-principle theoretical studies of hydrogen-rich compounds compressed to megabar level pressure, there is a very limited experimental dataset available for such materials. In this paper, we analyze the upper critical field, $B_{c2}(T)$, data of highly compressed H$_3$S reported by Mozaffari and co-workers (S. Mozaffari, et. al., *LA-UR*-18-30460, DOI: 10.2172/1481108) by utilizing four different models of $B_{c2}(T)$. In result, we find that the ratio of superconducting energy gap, $\Delta(0)$, to the Fermi energy, $\varepsilon_F$, in all considered scenarios is $0.03 < \Delta(0)/\varepsilon_F < 0.07$, with respective ratio of $T_c$ to the Fermi temperature, $T_F$, $0.012 < T_c/T_F < 0.039$. These characterize H$_3$S as unconventional superconductor and places it on the same trend line in $T_c$ versus $T_F$ plot, where all unconventional superconductors located.

# Classifying superconductivity in compressed H₃S

## I. Introduction

Experimental discovery a superconductivity above $T$ = 200 K in highly compressed $H_3S$ by Drozdov et al [1] is one of the most fascinating confirmation of the Bardeen-Cooper-Schrieffer (BCS) theory [2] and the phonon-mediated pairing scenario which can sustain superconductivity at such high temperature [3,4]. Moreover, recent experimental results on another hydrogen-rich compound of $LaH_{10}$ [5,6], further showed that BCS electron-phonon pairing mechanism works at much higher temperatures, and highest observed in experiment superconducting transition temperature, $T_c$, for $LaH_{10}$ compound is $T_c$ = 260 K [6]. Historical aspects of the discovery, included the astonishing theoretical prediction of Ashcroft [7], and reviews of theoretical works in the field can be found elsewhere [8-13].

Most theoretical works [10,12,13-18] came to conclusion that $H_3S$ is strong coupled superconductor with BCS ratio:

$$\frac{2 \cdot \Delta(0)}{k_B \cdot T_c} = \alpha = 4.5 - 4.7 \qquad (1)$$

where $\Delta(0)$ is ground state of the superconducting energy gap, $k_B$ is the Boltzmann constant. In contrast to this, our analysis [19] of experimental self-field critical current density, $J_c(sf,T)$ (reported by Drozdov and co-workers in [1]), showed that the BCS ratio (Eq. 1) for $H_3S$ is more likely to be very close to the weak-coupling limit of 3.53, and we deduced value for $\Delta(0)$ = 28 meV [19,20], while many theoretical works came to predicted values in the range of $\Delta(0)$ = 40-45 meV. Modern spectroscopic techniques have been applied to $H_3S$ [21], which confirmed theoretically calculated energy spectrum for energies above 70 meV.

In this paper, we analyse recently released experimental upper critical field, $B_{c2}(T)$, data [22] for highly compressed $H_3S$ with the purpose to deduce the Fermi velocity, $v_F$, and Fermi energy, $\varepsilon_F$, for this material.

## II. Description of models

In the Ginzburg-Landau theory, the upper critical field is given by following expression:

$$B_{c2}(T) = \frac{\phi_0}{2\cdot\pi\cdot\xi^2(T)} \qquad (2)$$

where $\phi_0 = 2.07\cdot 10^{-15}$ Wb is flux quantum, and $\xi(T)$ is the coherence length. There is a well-known BCS expression [2]:

$$\xi(0) = \frac{\hbar\cdot v_F}{\pi\cdot\Delta(0)} \qquad (3)$$

where $\hbar = h/2\pi$ is reduced Planck constant, and $v_F$ is the Fermi velocity. Thus, from deduced $B_{c2}(0)$ and $T_c$ and assumed $\alpha$ (Eq. 1), one can calculate the Fermi velocity, $v_F$:

$$v_F = \frac{\pi}{2}\cdot\xi(0)\cdot\frac{\alpha\cdot k_B\cdot T_c}{\hbar}, \qquad (4)$$

the Fermi energy, $\varepsilon_F$:

$$\varepsilon_F = \frac{m^*_{eff}\cdot v_F^2}{2} \qquad (5)$$

where $m^*_{eff}$ is effective mass (for H$_3$S we used $m^*_{eff} = 2.76\, m_e$ [10]), and the Fermi temperature, $T_F$:

$$T_F = \frac{\varepsilon_F}{k_B} \qquad (6)$$

where $k_B$ is Boltzmann constant.

One of conventional models to analyse $B_{c2}(T)$ was given by Werthamer-Helfand-Hohenberg (WHH) [23,24]:

$$ln\left(\frac{T}{T_c(B=0)}\right) = \psi\left(\frac{1}{2}\right) - \psi\left(\frac{1}{2} + \frac{\hbar\cdot D\cdot B_{c2}(T)}{2\cdot\phi_0\cdot k_B\cdot T}\right) \qquad (6)$$

where $D$ is the diffusion constant of the normal conducting electrons/holes, with two free fitting parameters of $T_c(B=0)$ and $D$. Baumgartner *et al* [25] proposed simple and accurate analytical expression for $B_{c2}(T)$ within WHH model:

$$B_{c2}(T) = \frac{1}{0.693}\cdot\frac{\phi_0}{2\cdot\pi\cdot\xi^2(0)}\cdot\left(\left(1-\frac{T}{T_c}\right) - 0.153\cdot\left(1-\frac{T}{T_c}\right)^2 - 0.152\cdot\left(1-\frac{T}{T_c}\right)^4\right) \qquad (7)$$

where ξ(0) and $T_c \equiv T_c(B=0)$ are two free fitting parameters. We will designate this model as B-WHH.

In addition, there are several analytical expressions which are in a wide use too [26-28]. For instance, there are classical two-fluid Gorter-Casimir model [29]:

$$B_{c2}(T) = \frac{\phi_0}{2\cdot\pi\cdot\xi^2(0)} \cdot \left(1 - \left(\frac{T}{T_c}\right)^2\right) \tag{8}$$

and Jones-Hulm-Chandrasekhar (JHC) model [30]:

$$B_{c2}(T) = \frac{\phi_0}{2\cdot\pi\cdot\xi^2(0)} \cdot \left(\frac{1-\left(\frac{T}{T_c}\right)^2}{1+\left(\frac{T}{T_c}\right)^2}\right) \tag{9}$$

There is also a little-known equation from Gor'kov for $B_{c2}(T)$ [31] which was referred by Gor'kov as a good analytical interpolative approximation over the whole temperature range:

$$B_{c2}(T) = B_c(T) \cdot \frac{\sqrt{2}}{1.77} \cdot \frac{\lambda(0)}{\xi(0)} \cdot \left(1.77 - 0.43 \cdot \left(\frac{T}{T_c}\right)^2 + 0.07 \cdot \left(\frac{T}{T_c}\right)^4\right) \tag{10}$$

where $B_c(T)$ is the thermodynamic critical field, and λ(0) is the ground state London penetration depth. Eq. 8 was re-written by Jones *et al* [30] in following form:

$$B_{c2}(T) = \frac{1}{1.77} \cdot \frac{\phi_0}{2\cdot\pi\cdot\xi^2(0)} \cdot \left(1.77 - 0.43 \cdot \left(\frac{T}{T_c}\right)^2 + 0.07 \cdot \left(\frac{T}{T_c}\right)^4\right) \cdot \left(1 - \left(\frac{T}{T_c}\right)^2\right) \tag{11}$$

We will designate Eq. 9 as G model.

In this paper, we utilise Eq. 8 in a different way. If we take in account, the Ginzburg-Landau (GL) theory expressions:

$$B_{c2}(T) = \sqrt{2} \cdot \frac{\lambda(T)}{\xi(T)} \cdot B_c(T) \tag{12}$$

we can conclude that the Gor'kov's equation (Eq. 8) means that:

$$\kappa(T) = \frac{\lambda(T)}{\xi(T)} = \frac{1}{1.77} \cdot \frac{\lambda(0)}{\xi(0)} \cdot \left(1.77 - 0.43 \cdot \left(\frac{T}{T_c}\right)^2 + 0.07 \cdot \left(\frac{T}{T_c}\right)^4\right) \tag{13}$$

By utilising another GL theory expression:

$$B_{c2}(T) = 2 \cdot \left(\frac{\lambda(T)}{\xi(T)}\right)^2 \cdot \frac{B_{c1}(T)}{\ln(\kappa(T))+0.5} = \left(\frac{\lambda(T)}{\xi(T)}\right)^2 \cdot \frac{\phi_0}{2\cdot\pi\cdot\lambda^2(T)} = \left(\kappa(T)\right)^2 \cdot \frac{\phi_0}{2\cdot\pi\cdot\lambda^2(T)} \tag{14}$$

and BCS expression for λ(T) for *s*-wave superconductor:

$$\lambda(T) = \frac{\lambda(0)}{\sqrt{1 - \frac{1}{2 \cdot k_B \cdot T} \cdot \int_0^\infty \frac{d\varepsilon}{\cosh^2\left(\frac{\sqrt{\varepsilon^2 + \Delta^2(T)}}{2 \cdot k_B \cdot T}\right)}}} \tag{15}$$

where the temperature-dependent superconducting gap $\Delta(T)$ equation can be taken from Gross *et al* [32]:

$$\Delta(T) = \Delta(0) \cdot \tanh\left[\frac{\pi \cdot k_B \cdot T_C}{\Delta(0)} \cdot \sqrt{\eta \cdot \frac{\Delta C}{C} \cdot \left(\frac{T_C}{T} - 1\right)}\right] \tag{16}$$

where $\Delta C/C$ is the relative jump in electronic specific heat at $T_c$, and $\eta = 2/3$ for *s*-wave superconductors [32], one can obtain expression for the temperature dependent upper critical field:

$$B_{c2}(T) = \frac{\phi_0}{2 \cdot \pi \cdot \xi^2(0)} \cdot \left[\left(\frac{1.77 - 0.43 \cdot \left(\frac{T}{T_C}\right)^2 + 0.07 \cdot \left(\frac{T}{T_C}\right)^4}{1.77}\right)^2 \cdot \frac{1}{1 - \frac{1}{2 \cdot k_B \cdot T} \cdot \int_0^\infty \frac{d\varepsilon}{\cosh^2\left(\frac{\sqrt{\varepsilon^2 + \Delta^2(T)}}{2 \cdot k_B \cdot T}\right)}}\right] \tag{17}$$

Thus, four fundamental parameters of superconductor, i.e. $\xi(0)$, $\Delta(0)$, $\Delta C/C$ and $T_c$, can be deduced by fitting experimental $B_{c2}(T)$ data to Eq. 17. We need to clarify that $\xi(0)$ determines absolute value of $B_{c2}(0)$ amplitude, while $\Delta(0)$ and $\Delta C/C$ are deduced from the shape of $B_{c2}(T)$ curve (which is the part of Eq. 17 in square brackets).

In this paper we fit experimental $B_{c2}(T)$ data for compressed sulfur hydride to Eqs. 7, 9, and 11, 17 with the purpose to deduce/calculate fundamental superconducting parameters of this material.

**III. Results and Discussions**

Mozaffari et al [22] in their Fig. 1(a) defined two values for the upper critical field:

1. At the onset of superconductivity, which we will designate as $B_{c2}(T)$ (in accordance with Mozaffari et al [22] definition).
2. At zero-resistance point, which we will designate as $B_{c2,R=0}(T)$ for the clarity.

In Figs. 1-4 we show raw upper critical field data and data fits to four models:

Panel a: B-WHH model [24] (Eq. 7);

Panel b: JHC model [30] (Eq. 9);

Panel c: G model [31] (Eq. 11);

Panel d: this work model (Eq. 17).

In Figs. 1,2 we show results for Sample #1 compressed at $P$ = 150 GPa. In Figs. 3,4 we show results for Sample #2 compressed at $P$ = 170 GPa. In Figs. 1,3 we analysed $B_{c2,R=0}(T)$ data, and in Figs. 2,4 we analysed $B_{c2}(T)$ data. Results of all fits are presented in Table 1.

In general (Figs. 1-4, Table 1), we can conclude that all four models provide good fit quality, $R$, and deduced values of $T_c$ and $\xi(0)$ for all four models are in reasonable agreement with each other. The most interesting thing we found is that fits to Eq. 17 reveal for all four $B_{c2}(T)$ datasets the value for superconducting energy gap of $\Delta(0)$ = 25-28 meV which all are in excellent agreement with the value we deduced by the analysis of critical current densities in $H_3S$ in our previous work [19], $\Delta(0)$ = 28 meV. The latter was deduced for different $H_3S$ sample [1] with $T_c$ = 203 K, while in present work we analysed data for samples with lower $T_c$.

All deduced $B_{c2}(0)$ values (Fig. 1-4) are well below Pauli limit of:

$$B_p(0) = \frac{2 \cdot \Delta(0)}{g \cdot \mu_B} = 430 - 500 \, T \gg B_{c2}(0) \tag{18}$$

where $g$ = 2 and $\mu_B = \frac{e \cdot \hbar}{2 \cdot m_e}$ is the Bohr magneton. Following Gor'kov's note [33], Eq. 18 means that the mean-free path, $l$, of the electrons is large compared with the coherence length:

$$l \gg \xi(T) > \xi(0) \sim 2.5 \, nm \tag{19}$$

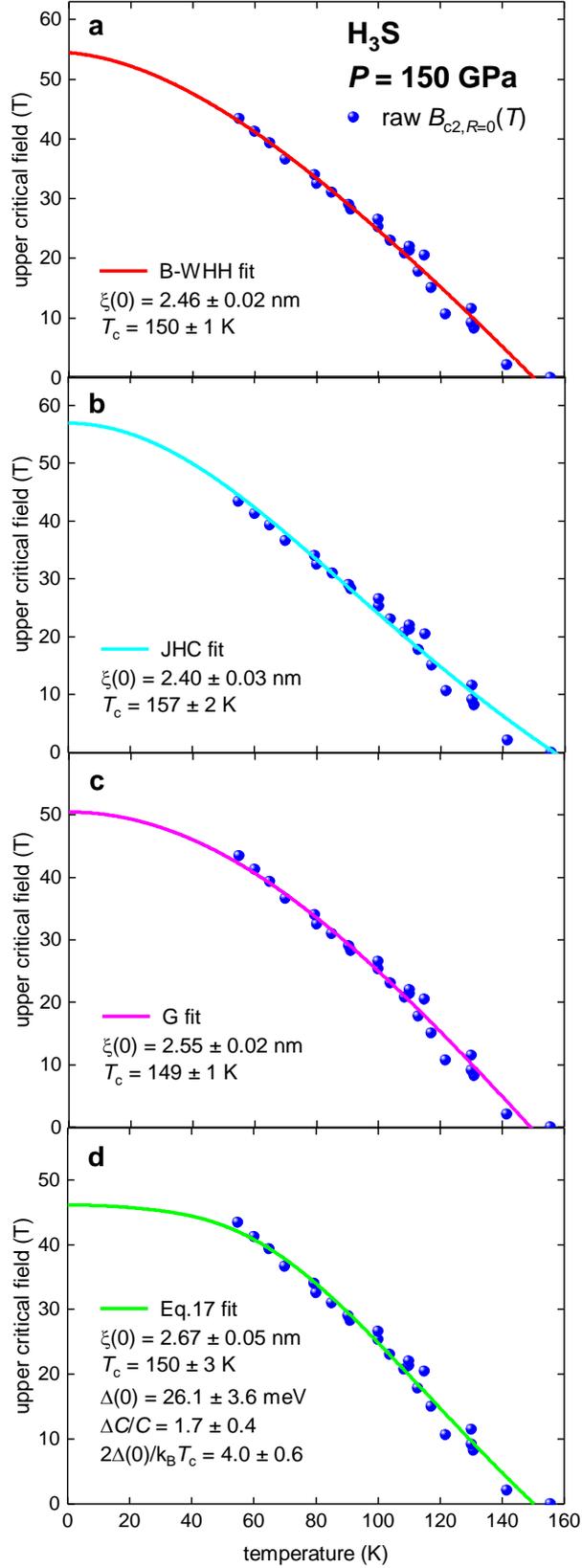

**Figure 1.** Superconducting upper critical field, $B_{c2,R=0}(T)$, data (blue) for compressed $H_3S$ Sample #1 at pressure $P = 150$ GPa (raw data are from Ref. 22). (a) Fit to B-WHH model [24] (Eq. 7), fit quality is $R = 0.9832$. (b) Fit to JHC model [30] (Eq. 9), fit quality is $R = 0.9785$. (c) Fit to G model [31] (Eq. 11), fit quality is $R = 0.9827$. (d) Fit to this work model (Eq. 17), fit quality is $R = 0.9832$.

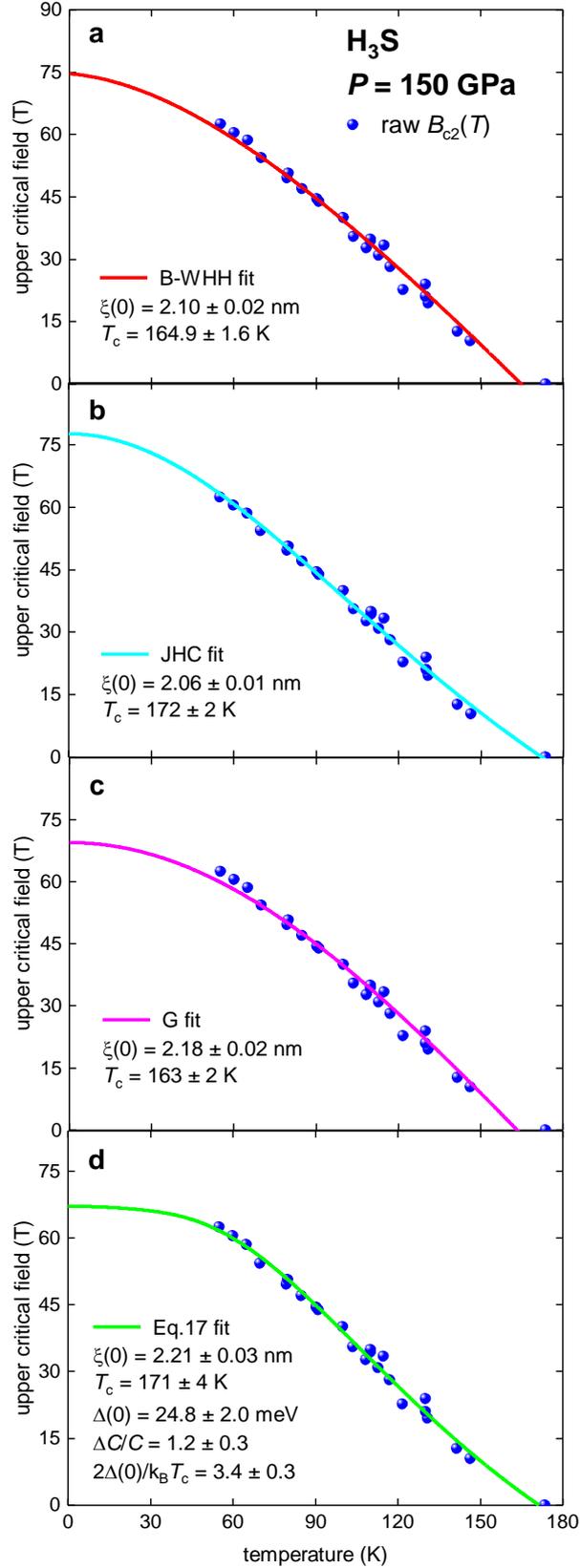

**Figure 2.** Superconducting upper critical field, $B_{c2}(T)$, data (blue) for compressed H$_3$S Sample #1 at pressure $P = 150$ GPa (raw data are from Ref. 22). (a) Fit to B-WHH model [24] (Eq. 7), fit quality is $R = 0.9850$. (b) Fit to JHC model [30] (Eq. 9), fit quality is $R = 0.9908$. (c) Fit to G model [31] (Eq. 11), fit quality is $R = 0.9806$. (d) Fit to this work model (Eq. 17); fit quality is $R = 0.9914$.

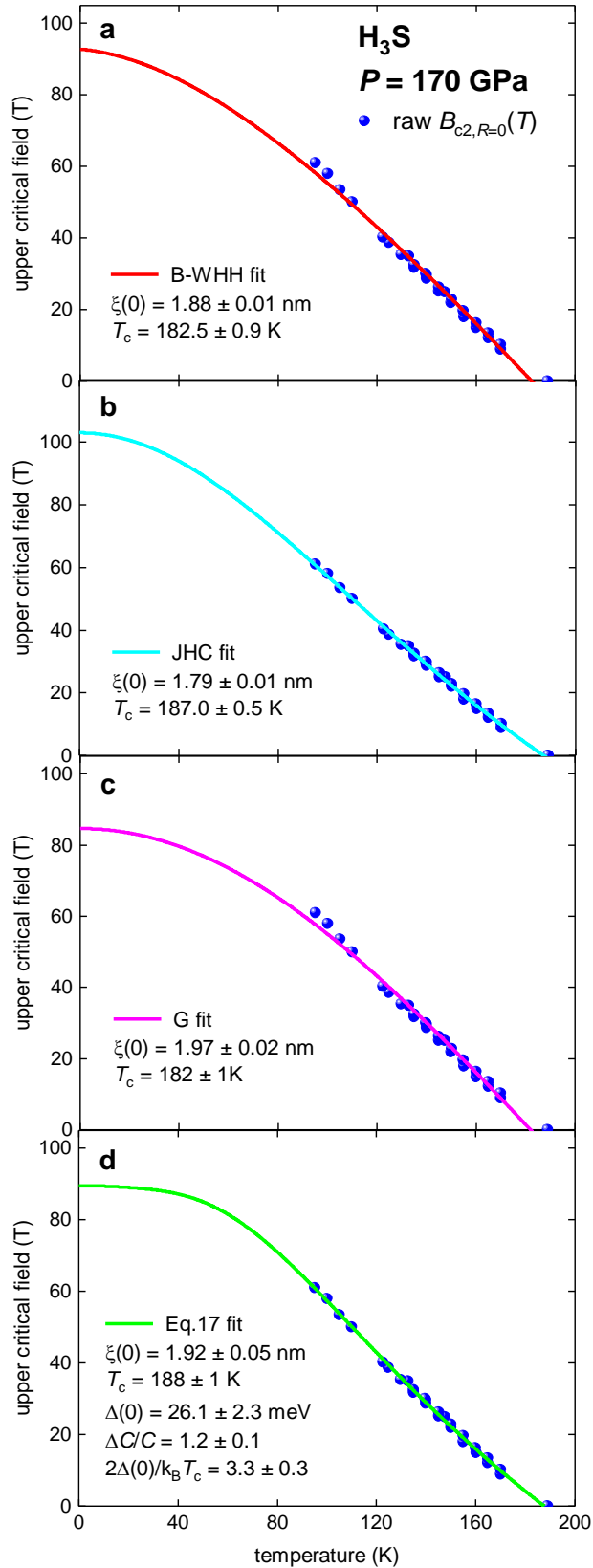

**Figure 3.** Superconducting upper critical field, $B_{c2,R=0}(T)$, data (blue) for compressed H$_3$S Sample #2 at pressure $P = 170$ GPa (raw data are from Ref. 22). (a) Fit to B-WHH model [24] (Eq. 7), fit quality is $R = 0.9901$. (b) Fit to JHC model [30] (Eq. 9), fit quality is $R = 0.9978$. (c) Fit to G model [31] (Eq. 11), fit quality is $R = 0.9879$. (d) Fit to this work model (Eq. 17); fit quality is $R = 0.9979$.

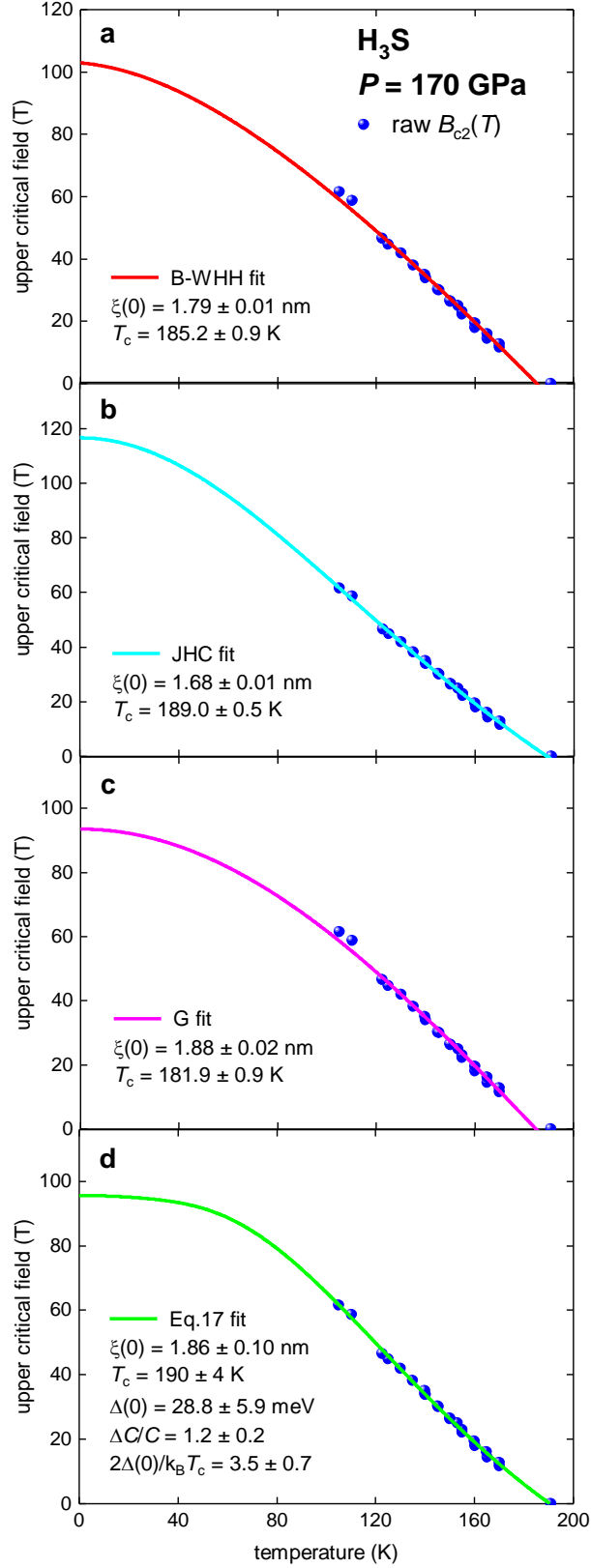

**Figure 4.** Superconducting upper critical field, $B_{c2}(T)$, data (blue) for compressed $H_3S$ Sample #2 at pressure $P$ = 170 GPa (raw data are from Ref. 22). (a) Fit to B-WHH model [24] (Eq. 7), fit quality is $R$ = 0.990. (b) Fit to JHC model [30] (Eq. 9), fit quality is $R$ = 0.9978. (c) Fit to G model [31] (Eq. 11), fit quality is $R$ = 0.9886. (d) Fit to this work model (Eq. 17); fit quality is $R$ = 0.9981.

This is interesting result, if we take in account that $H_3S$ is formed by chemical reaction which occurs within the diamond anvil volume:

$$3H_2S \rightarrow 2H_3S + S \qquad (20)$$

and pure sulfur is always presented as post-reacted product in the studied sample.

However, Eq. 18 tells us that two phases, i.e. $H_3S$ and S, are reasonably well separated from each other and there is a very low level of atomic disordering within superconducting $H_3S$ phase, which has lattice parameter of $a = 0.3092$ nm [34].

The next step of the analysis is the comparison of $v_F$, $\varepsilon_F$, $T_F$ values calculated directly by Eq. 3 (because fits to Eq. 17 provide both required quantities, i.e. $\xi(0)$ and $\Delta(0)$) with $v_F$ values calculated by Eq. 4 in assumption of two extreme coupling strength scenario of $\alpha = 3.53$ and $\alpha = 4.70$. Overall, deduced/calculated $v_F$ for $H_3S$ are in the range of $v_F = (2.0\text{-}3.8) \times 10^5$ m/s which equals to $v_F$ of nickel and cobalt at normal conditions [35] and is approximately equal to the universal nodal Fermi velocity of the superconducting cuprates [36].

**Table 1.** Deduced parameters for $H_3S$ superconductor. We assumed that electron effective mass in $H_3S$ is $m_{eff} = 2.76\, m_e$ [10].

| Pressure (GPa) | Raw data | Model | Deduced $T_c$ (K) | Deduced $\xi(0)$ (nm) | Assumed/deduced $\frac{2\cdot\Delta(0)}{k_B \cdot T_c}$ | $\Delta C/C$ | $v_F$ ($10^5$ m/s) | $\Delta(0)$ meV | $\varepsilon_F$ eV | $\Delta(0)/\varepsilon_F$ | $T_F$ ($10^3$ K) | $T_c/T_F$ |
|---|---|---|---|---|---|---|---|---|---|---|---|---|
| 150 | $B_{c2,R=0}$ (T) | B-WHH | 150 ± 1 | 2.46 ± 0.02 | 3.53 | | 2.68 ± 0.03 | 22.8 ± 0.2 | 0.56 ± 0.01 | 0.040 ± 0.001 | 6.5 ± 0.2 | 0.023 ± 0.001 |
| | | | | | 4.70 | | 3.57 ± 0.04 | 30.4 ± 0.04 | 1.00 ± 0.02 | 0.030 ± 0.001 | 11.6 ± 0.4 | 0.013 ± 0.001 |
| | | JHC | 157 ± 2 | 2.40 ± 0.03 | 3.53 | | 2.74 ± 0.03 | 23.9 ± 0.4 | 0.59 ± 0.03 | 0.041 ± 0.002 | 6.8 ± 0.2 | 0.023 ± 0.001 |
| | | | | | 4.70 | | 3.65 ± 0.05 | 31.8 ± 0.4 | 1.04 ± 0.02 | 0.030 ± 0.002 | 12.1 ± 0.5 | 0.013 ± 0.001 |
| | | G | 149 ± 1 | 2.55 ± 0.02 | 3.53 | | 2.76 ± 0.03 | 22.7 ± 0.2 | 0.60 ± 0.01 | 0.038 ± 0.002 | 6.9 ± 0.2 | 0.021 ± 0.001 |
| | | | | | 4.70 | | 3.68 ± 0.04 | 30.2 ± 0.3 | 1.06 ± 0.02 | 0.028 ± 0.001 | 12.3 ± 0.5 | 0.012 ± 0.001 |
| | | Eq. 16 | 150 ± 3 | 2.67 ± 0.05 | 4.0 ± 0.6 | 1.7 ± 0.4 | 3.33 ± 0.45 | 26.1 ± 3.6 | 0.87 ± 0.12 | 0.030 ± 0.004 | 10.1 ± 1.4 | 0.015 ± 0.002 |
| | | | | | 3.53 | | 2.51 ± 0.03 | 25.1 ± 0.3 | 0.50 ± 0.01 | 0.051 ± 0.002 | 5.8 ± 0.2 | 0.029 ± 0.001 |

| | | | | | | | | | | | |
|---|---|---|---|---|---|---|---|---|---|---|---|
| | | B-WHH | 165 ± 2 | 2.10 ± 0.02 | 4.70 | | 3.35 ± 0.03 | 33.4 ± 0.3 | 0.88 ± 0.02 | 0.038 ± 0.002 | 10.2 ± 0.2 | 0.016 ± 0.001 |
| | | JHC | 172 ± 2 | 2.06 ± 0.01 | 3.53 | | 2.57 ± 0.03 | 26.2 ± 0.3 | 0.52 ± 0.01 | 0.050 ± 0.002 | 6.0 ± 0.2 | 0.029 ± 0.001 |
| | | | | | 4.70 | | 3.42 ± 0.03 | 34.8 ± 0.3 | 0.92 ± 0.02 | 0.038 ± 0.001 | 10.7 ± 0.2 | 0.016 ± 0.001 |
| | $B_{c2}(T)$ | G | 163 ± 2 | 2.18 ± 0.02 | 3.53 | | 2.58 ± 0.05 | 24.8 ± 0.3 | 0.52 ± 0.02 | 0.048 ± 0.002 | 6.0 ± 0.2 | 0.027 ± 0.001 |
| | | | | | 4.70 | | 3.43 ± 0.07 | 33.0 ± 0.3 | 0.92 ± 0.05 | 0.036 ± 0.002 | 10.7 ± 0.2 | 0.015 ± 0.001 |
| | | Eq. 16 | 171 ± 4 | 2.21 ± 0.03 | 3.4 ± 0.3 | 1.2 ± 0.3 | 2.6 ± 0.3 | 24.8 ± 2.0 | 0.54 ± 0.06 | 0.046 ± 0.005 | 6.3 ± 0.5 | 0.027 ± 0.003 |
| 170 | $B_{c2,R=0}(T)$ | B-WHH | 182.5 ± 0.9 | 1.88 ± 0.01 | 3.53 | | 2.50 ± 0.03 | 27.8 ± 0.1 | 0.49 ± 0.01 | 0.057 ± 0.002 | 5.7 ± 0.2 | 0.032 ± 0.002 |
| | | | | | 4.70 | | 3.32 ± 0.02 | 37.0 ± 0.2 | 0.87 ± 0.01 | 0.043 ± 0.001 | 10.0 ± 0.2 | 0.018 ± 0.001 |
| | | JHC | 187.0 ± 0.5 | 1.79 ± 0.01 | 3.53 | | 2.43 ± 0.01 | 28.4 ± 0.1 | 0.46 ± 0.01 | 0.062 ± 0.001 | 5.4 ± 0.1 | 0.035 ± 0.001 |
| | | | | | 4.70 | | 3.23 ± 0.01 | 37.9 ± 0.1 | 0.82 ± 0.01 | 0.046 ± 0.001 | 9.5 ± 0.1 | 0.020 ± 0.001 |
| | | G | 182 ± 1 | 1.97 ± 0.02 | 3.53 | | 2.61 ± 0.01 | 27.7 ± 0.1 | 0.53 ± 0.01 | 0.052 ± 0.001 | 6.2 ± 0.1 | 0.030 ± 0.001 |
| | | | | | 4.70 | | 3.47 ± 0.01 | 36.9 ± 0.1 | 0.94 ± 0.01 | 0.052 ± 0.001 | 11.0 ± 0.2 | 0.017 ± 0.001 |
| | | Eq. 16 | 188 ± 1 | 1.92 ± 0.05 | 3.3 ± 0.3 | 1.2 ± 0.1 | 2.4 ± 0.2 | 26.1 ± 2.3 | 0.44 ± 0.05 | 0.059 ± 0.006 | 5.0 ± 0.5 | 0.037 ± 0.004 |
| | $B_{c2}(T)$ | B-WHH | 185.2 ± 0.9 | 1.79 ± 0.01 | 3.53 | | 2.40 ± 0.01 | 28.2 ± 0.1 | 0.45 ± 0.01 | 0.062 ± 0.001 | 5.3 ± 0.1 | 0.035 ± 0.001 |
| | | | | | 4.70 | | 3.20 ± 0.01 | 37.5 ± 0.1 | 0.80 ± 0.02 | 0.047 ± 0.002 | 9.3 ± 0.2 | 0.020 ± 0.001 |
| | | JHC | 189.0 ± 0.5 | 1.68 ± 0.01 | 3.53 | | 2.30 ± 0.01 | 28.7 ± 0.1 | 0.42 ± 0.01 | 0.069 ± 0.002 | 4.8 ± 0.2 | 0.039 ± 0.001 |
| | | | | | 4.70 | | 2.07 ± 0.01 | 38.3 ± 0.1 | 0.74 ± 0.02 | 0.052 ± 0.002 | 8.6 ± 0.2 | 0.022 ± 0.001 |
| | | G | 181.9 ± 0.9 | 1.97 ± 0.02 | 3.53 | | 2.48 ± 0.02 | 27.7 ± 0.1 | 0.48 ± 0.01 | 0.057 ± 0.002 | 5.6 ± 0.3 | 0.033 ± 0.002 |
| | | | | | 4.70 | | 3.30 ± 0.02 | 36.8 ± 0.1 | 0.85 ± 0.02 | 0.043 ± 0.002 | 9.9 ± 0.3 | 0.018 ± 0.001 |
| | | Eq. 16 | 190 ± 4 | 1.86 ± 0.10 | 3.5 ± 0.7 | 1.2 ± 0.2 | 2.6 ± 0.4 | 28.8 ± 5.9 | 0.51 ± 0.09 | 0.056 ± 0.009 | 6.0 ± 1.0 | 0.032 ± 0.006 |

Examination of the values in Table I leaded us to three important findings:

1. The ratio of the superconducting energy gap, $\Delta(0)$, to the Fermi energy, $\varepsilon_F$, in all considered scenarios (including direct deduction by Eq. 17) is within interval of $0.03 < \Delta(0)/\varepsilon_F < 0.07$. These values characterize H$_3$S material as an unconventional superconductor, by illustration, conventional niobium, Nb, has the ratio which is at least two orders of magnitude lower, i.e. $\Delta(0)/\varepsilon_F = 3 \cdot 10^{-4}$ [37].

2. The most straightforward way to see our conclusion that H$_3$S is unconventional superconductor is to add $T_c$ and $T_F$ data for H$_3$S on the plot of $T_c$ versus $T_F$ where other superconductors are shown. In this plot (Fig. 5) (data in Fig. 5 were adopted from Uemura [38], Ye et al [39], Qian et al [40], and Hashimoto et al [41]) all unconventional superconductors are located within a narrow band of $0.01 < T_c/T_F < 0.05$. We note that Uemura [38] stated that there is the upper limit for $T_c/T_F = 0.05$ for all known superconductors. In all considered scenarios, $H_3S$ has ratios within interval of $0.012 < T_c/T_F < 0.039$ (Fig. 5 and Table 1). It is clearly visible in Fig. 5 that $H_3S$ is in the same band where all unconventional superconductors, particularly heavy fermions and cuprates, are. In this regard, $H_3S$ is located just above Bi-2223 phase. In this regard, $H_3S$ is the material which is located at the position where majority of others unconventional superconductors placed.

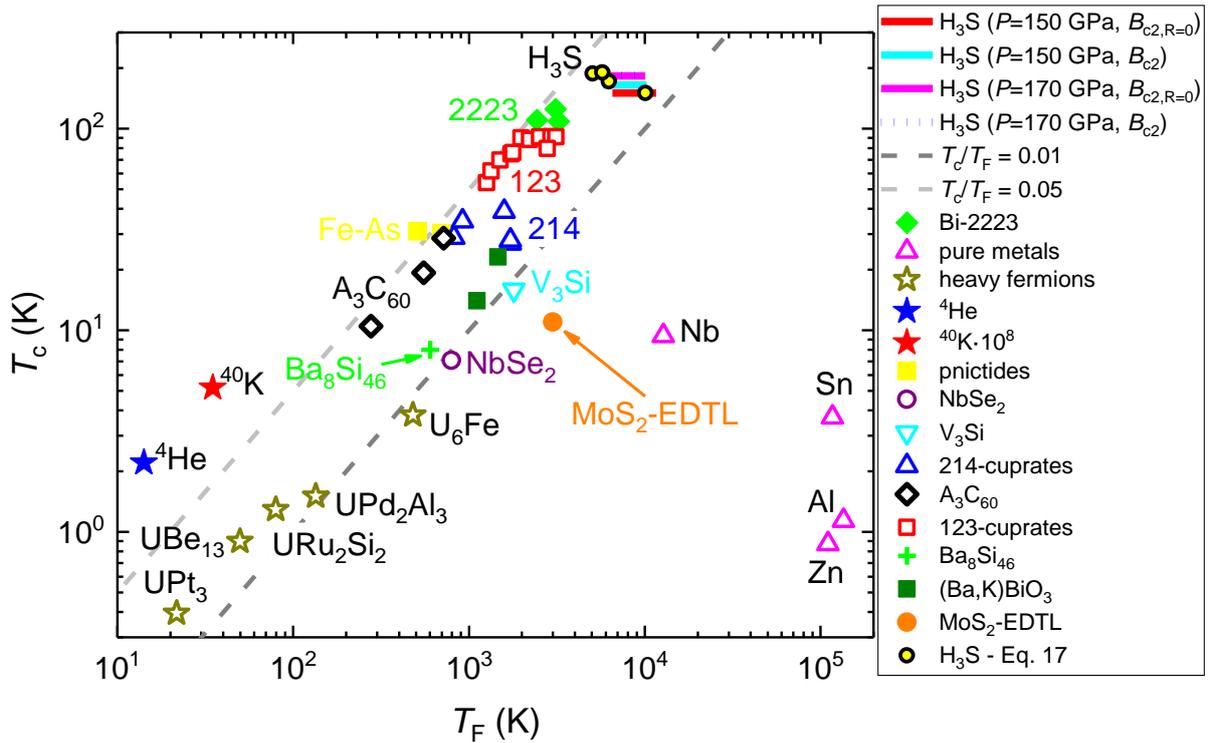

**Figure 5.** A plot of $T_c$ versus $T_F$ obtained for most representative superconducting families. Data was taken from Uemura [38], Ye et al [39], Qian et al [40], and Hashimoto et al [41].

3. We also can see that despite of very different assumptions and varieties of the upper critical field data definition, the Fermi velocity is within reasonably narrow interval of $v_F = (2.1-3.7) \cdot 10^5$ m/s. This value is about two times lower than $v_F$ of alkali metals at normal conditions [35,37] and it approximately equals to the universal nodal Fermi velocity of the superconducting cuprates [36]. This is another manifestation that H<sub>3</sub>S should be classified as unconventional superconductor.

Even though the original paper from Drozdov et. al. [1] stated that H<sub>3</sub>S is conventional superconductor, and this point of view was very quickly widely accepted by the scientific community [3], we must note that at that time there were no available experimental data which supported this point of view. One of prerequisites of phonon mediated mechanism in H<sub>3</sub>S is the strong-coupling electron-phonon interaction (references on original papers can be found in Ref. 13), which we cannot confirm neither by the analysis of experimental critical current densities [20], nor by the analysis of experimental upper critical field data presented herein. Instead our analysis gives clear evidence that H<sub>3</sub>S is weak-coupled superconductor, with the ratio:

$$3.3 \pm 0.3 < \frac{2 \cdot \Delta(0)}{k_B \cdot T_c} < 4.0 \pm 0.6 \qquad (21)$$

and average value of

$$\frac{2 \cdot \Delta(0)}{k_B \cdot T_c} = 3.55 \pm 0.31 \qquad (22)$$

which is remarkably closed to weak-coupling limit of BCS theory of 3.53. Average absolute value of the ground state superconducting energy gap is:

$$\Delta(0) = 26.5 \pm 1.7 \, meV \qquad (23)$$

This value is in a very good agreement with Δ(0) = 27.8 meV which we deduced in our previous paper by the analysis of critical current density in H<sub>3</sub>S [19] for sample with $T_c$ = 203 K.

**IV. Conclusion**

In this paper, we analysed the upper critical field data for compressed H<sub>3</sub>S which were recently released by Los-Alamos Laboratory [22]. Result of our analysis showed that compressed H$_3$S should be classified as another member of unconventional superconductor family.

**Acknowledgement**

Author thanks Ratu Mataira-Cole (University of Wellington, New Zealand) for reading, commenting and editing the manuscript. Author thanks financial support provided by the state assignment of FASO of Russia (theme "Pressure" No. AAAA-A18-118020190104-3) and by Act 211 Government of the Russian Federation, contract No. 02.A03.21.0006.

**References**

[1]  A.P. Drozdov, M. I. Eremets, I. A. Troyan, V. Ksenofontov, S. I. Shylin, *Nature* **525**, 73 (2015).

[2]  J. Bardeen, L. N. Cooper, J. R. Schrieffer *Phys. Rev.* **108**, 1175 (1957).

[3]  I. I. Mazin, *Nature* **525**, 40 (2015).

[4]  N. Bernstein, C. S. Hellberg, M. D. Johannes, I. I. Mazin, M. J. Mehl, *Physical Review B* **91**, 060511 (2015).

[5]  A. P. Drozdov, V. S. Minkov, S. P. Besedin, P. P. Kong, M A. Kuzovnikov, D. A. Knyazev, M I. Eremets, arXiv:1808.07039 (2018).

[6]  M. Somayazulu, M. Ahart, A. K. Mishra, Z. M. Geballe, M. Baldini, Y. Meng, V. V. Struzhkin, R. J. Hemley, arXiv:1808.07695 (2018).

[7]  N. W. Ashcroft, *Phys. Rev. Lett.* **21**, 1748 (1968).

[8]  M. I. Eremets and A. P. Drozdov, *Phys.-Usp.* **59**, 1154 (2016).

[9]  I. Troyan, et. al., *Science* **351**, 1303 (2016).

[10]  A. P. Durajski, *Sci. Rep.* **6**, 38570 (2016).

[11]  M. Einaga, et. al. *Japanese Journal of Applied Physics* **56**, 05FA13 (2017).

[12] H. Liu, I. I. Naumov, R. Hoffmann, N. W. Ashcroft, R. J. Hemley, *Proc. Natl. Acad. Sci. U. S. A.* **114**, 6990 (2017).

[13] L. P. Gor'kov, V. Z. Kresin, *Rev. Mod. Phys.* **90**, 011001 (2018).

[14] E. J. Nicol, J. P. Carbotte, *Phys. Rev. B* **91**, 220507(R) (2015).

[15] I. Errea, et. al., *Phys. Rev. Lett.* **114**, 157004 (2015).

[16] T. Jarlborg, A. Bianconi, *Scientific Reports* **6**, 24816 (2016).

[17] A. P. Durajski, R. Szczesniak, arXiv:1609.06079 (2016).

[18] A. P. Durajski, R. Szczęśniak, L. Pietronero, *Annalen der Physik* **528**, 358 (2016).

[19] E. F. Talantsev, W. P. Crump, J. G. Storey, J. L. Tallon *Annalen der Physik* **529**, 1600390 (2017).

[20] E. F. Talantsev, W. P. Crump, J. L. Tallon, *Annalen der Physik* **529**, 1700197 (2017).

[21] F. Capitani, et. al., *Nature Physics* **13**, 859 (2017).

[22] S. Mozaffari, et. al., *Los Alamos National Laboratory*, Report LA-UR-18-30460 (2018); DOI: 10.2172/1481108.

[23] E. Helfand, N. R. Werthamer, *Phys. Rev.* **147** 288 (1966).

[24] N. R. Werthamer, E. Helfand and P. C. Hohenberg, *Phys. Rev.* **147**, 295 (1966).

[25] T. Baumgartner, M. Eisterer, H. W. Weber, R. Fluekiger, C. Scheuerlein, L. Bottura, *Supercond. Sci. Technol.* **27**, 015005 (2014).

[26] F. Yuan, et. al. *New Journal of Physics* **20**, 093012 (2018).

[27] B. Pal, et. al. *Supercond. Sci. Technol.* **32** 015009 (2019).

[28] E. F. Talantsev, W. P. Crump, J. O. Island, Y. Xing, Y. Sun, J. Wang, J. L. Tallon *2D Materials* **4**, 025072 (2017).

[29] C. J. Gorter, H. Casimir, On supraconductivity I *Physica* **1**, 306-320 (1934).

[30] C. K. Jones, J. K. Hulm, B. S. Chandrasekhar, *Rev. Mod. Phys.* **36**, 74 (1964).

[31] L. P. Gor'kov, *Soviet Physics JETP* **10**, 593 (1960).

[32] F. Gross, B. S. Chandrasekhar, D. Einzel, K. Andres, P. J. Hirschfeld, H. R. Ott, J. Beuers, Z. Fisk, J. L. Smith, *Z. Phys. B - Condensed Matter* **64**, 175 (1986).

[33] L. P. Gor'kov, *Soviet Physics JETP* **17**, 518 (1963).

[34] M. Einaga, et. al., *Nature Physics* **12**, 835 (2016).

[35] D. Gall, *J. Appl. Phys.* **119**, 085101 (2016).

[36] X. J. Zhou, *Nature* **423**, 398 (2003).

[37] N. W. Aschroft, N. D. Mermin, Solid State Physics (Harcourt College Publishing, 1976) ISBN: 0030839939.

[38] Y. J. Uemura, *J. Phys.: Condens. Matter* **16** S4515 (2004).

[39] J. T. Ye, Y. J. Zhang, R. Akashi, M. S. Bahramy, R. Arita, Y. Iwasa, *Science* **338**, 1193 (2012).

[40] T. Qian, et. al., *Phys. Rev. Lett.* **106**, 187001 (2001).

[41] K. Hashimoto, et. al., *Science* **336**, 1554 (2012).